# What does Hirsch index evolution explain us?
# A case study: Turkish Journal of Chemistry

Metin Orbay, Orhan Karamustafaoğlu and Feda Öner
Amasya University (Turkey)
morbay@omu.edu.tr, orseka@yahoo.com, oner3@yahoo.com

Abstract

The evolution of Turkish Journal of Chemistry's (TURK J CHEM) Hirsch index (h-index) over the period 1995-2005 is studied and determined in the case of the self and without self-citations. It is seen that the effect of Hirsch index of TURK J CHEM has a highly positive trend during the last five years. It proves that TURK J CHEM is improving itself both in quantity and quality since h-index reflects peer review, and peer review reflects research quality of a journal.

Resumen

Presenta la evolución del índice Hirsch (h-index) de la Turkish Journal of Chemistry's (TURK J CHEM) durante el período 1995-2005; el caso es estudiado y precisado tanto con auto-citaciones como sin auto-citaciones. El indice Hirsch permite apreciar una alta tendencia positiva de los indicadores de la TURK J CHEM durante los últimos cinco años. Se comprueba así que la TURK J CHEM va mejorando tanto en cantidad como en calidad de contenidos dado que el h-index refleja la revisión por pares, y la revisión por pares refleja la calidad de las investigaciones de una revista.

Palabras clave

Citas bibliográficas, índice de citación, indice-h, NúmeroHirsch, Cienciometría, Análisis cienciométrico, Bibliometría

Introduction

Hitherto, several citation-based indicators have been used to measure research performance (e.g. the number of citations to each of the q most cited papers, the total number of citations, the citations per paper, the number of highly cited published papers). There are valid reservations about using above mentioned indicators to measure performance because some papers are cited for reasons that are unrelated to the quality or utility of a study (see: Kelly & Jennions, 2006; Miller, 2007 and references therein).

Recently, taking into account above citation-based indicators with advantages and disadvantages, Jorge E. Hirsch has suggested a new indicator called h-index, which means that one single index for the assessment of the research performance of an individual scientist. According to the definition by Hirsch, "A scientist has index h if his/her N papers have at least h citations each, and the other (N-h) papers have fewer than h citations each" (Hirsch, 2005). Hirsch's article has generated considerable interest and almost immediately provoked reactions in the scientific community (Ball, 2005; Braun, Glanzel & Schubert, 2005; Glanzel & Persson, 2005; Glanzel, 2006a; Egghe & Rousseau, 2006; Egghe, 2006; Cronin & Meho, 2006; Burrell, 2007; Rousseau, 2007a). The h-index has generally well received by the research group. Of course, the h-index has also a number of disadvantages as point out by some authors (Kelly & Jennions, 2006; Van Raan, 2006). After all these beneficial arguments, W. Glanzel has summarized some pros and cons of h-index in his excellent recent paper (Glanzel, 2006b).

After a short time, the h-index definition has been adapted into journals and article citations, as a h-type index-equal to h if you have published h papers, each of which has at least h citations (Braun, Glanzel & Schubert, 2006). T. Braun and co-workers stressed that the h-type index for journals would advantageously supplement journal impact factor (IF), the total number of citations divided by the number of articles (Garfield, 1955), at least two aspects: respectively,

i. It is robust in the sense that it is insensitive to an accidental excess of uncited articles, and to one or several highly cited articles,
ii. It combines the effect of "quantity" and "quality" in a rather specific.

Naturally, the journal h-index would not be calculated for a "lifetime contributions", as defined by Hirsch for the scientific output of a researcher, but for a definite period-in the simplest case for a given year. Using this procedure, R. Rousseau studied the evolution of the Journal of American Society of Information Sciences' Hirsch index and introduced relative h-index (Rousseau, 2007b).

In this opinion article, the evolution of Hirsch index of Turkish Journal of Chemistry (TURK J CHEM) over period 1995-2005 is studied and determined in the case of the self and elimination self-citation (or without self citation) of the journal.

Method and results

As is well known, Web of Science database offers a very simple way to determine the annual h-index of a journal. Retrieving all source items of a given journal from a given period and shorting them by the number of "times cited", it is easy to find the h-index of the journal for the given year (http://isiknowledge.com).

In this study, we conduct a case study for h-index of TURK J CHEM over period 1995-2005. Meanwhile, we consider a fixed moment in time when citations are collected from Web of Science (http://isiknowledge.com, retrieved date 24.12.2006). The h-index of TURK J CHEM over the period 1995-2005 is determined in the case of the self and without self-citations, as shown in Figure 1.

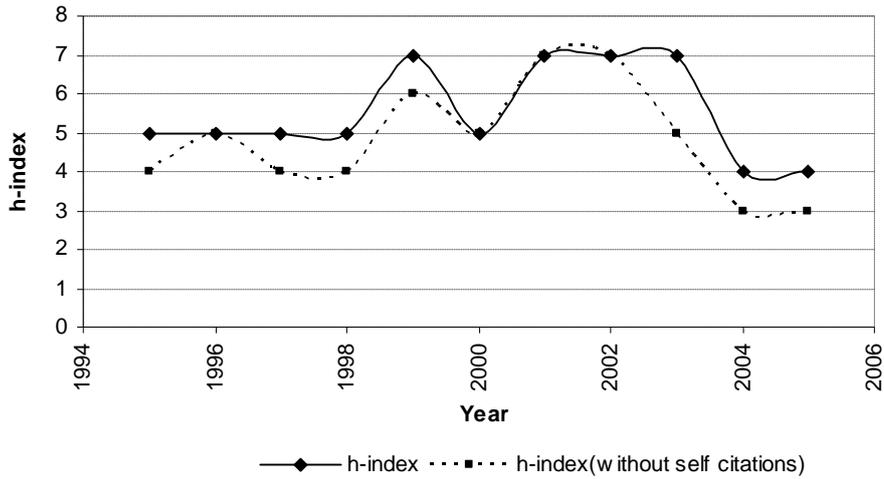

**Figure 1**. *h*-index of TURK J CHEM.

However, besides the period over which a volume can collect citations, also the number of articles published in that volume influences the h-index. For this reason, the h-index must be divided by the number of articles published, leading to a normalized (or relative) h-index (Rousseau, 2007b). In this case, the results are shown with self and without self citations for the journal in Figure 2.a. and Figure 2.b., respectively.

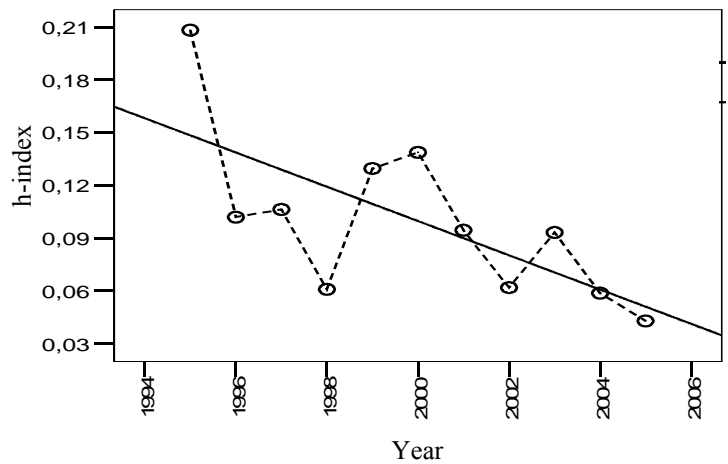

**Figure 2a.** Normalized *h* index with self citiations of TURK J CHEM.

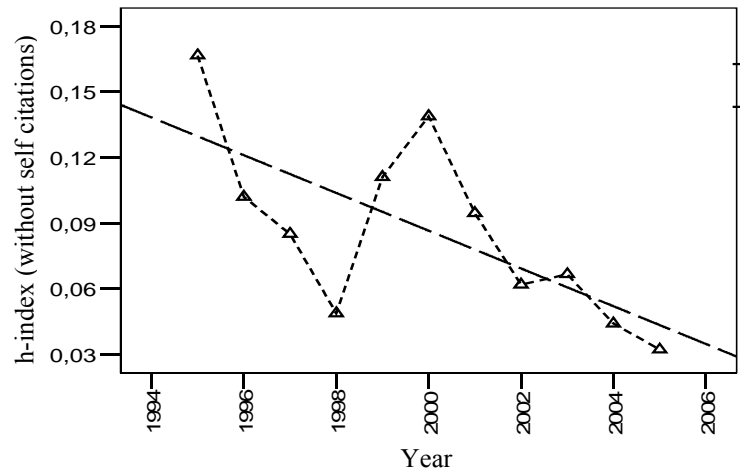

**Figure 2b.** Normalized *h* index with eliminated self citations of TURK J CHEM.

As can be clearly seen from Figure 2, using the normalized h-index leads to a linear increase when going backward in time (or decrease when going forward in time). The Pearson correlation coefficients of the regression lines of this journal are 0.686 for normalized h-index (continuous line in Figure 2.a) with self citations and 0.689 for normalized h-index without self-citations (dot line in Figure 2.b), which are moderate, but statistically significant (1% level). It is not surprising that these two correlation coefficients are very close to each other because of the fact that the self-citations over this period are limited by approximately 20%. On the other hand, we encounter that this value is high in other twenty randomly selected journals published in the same field.

It is obvious from the Figure 1 and Figure 2 that h-index and normalized h-index are extremely different trend between 1995-2000 and 2000-2005 periods. So, we focus on two periods. In the former period, the Pearson correlation coefficients are 0.336 for normalized h-index with self citations and 0.192 for without self-citation, which are very low, but statistically significant (1% level). It is not surprising that TURK J CHEM has started to be scanned in Web of Science newly in this period. For this reason, it can be thought that a few researchers were aware of this journal. On the other hand, in the latter period, the Pearson correlation coefficients are 0.858 for normalized h-index with self citations and 0.941 for without self-citation, which are very high, and statistically significant (1% level). From these interesting results, we conclude that a lot of published papers in this journal have been very high impact with respect to "quantity" (number of publications) and "quality" (citation rate), recently. Furthermore, it is known that TURK J CHEM has started open access in the latter period. Thus, we think that open access contributes h-index of this journal.